\documentclass[10pt, conference, compsocconf]{IEEEtran}

\usepackage{cite}
\usepackage[pdftex]{graphicx}
\usepackage{amsmath}
\usepackage{algorithm}
\usepackage[noend]{algpseudocode}
\usepackage{balance}
\usepackage{subfig}
\usepackage{tabularx}
\usepackage{multirow}
\usepackage{xcolor}
\usepackage{enumitem}
\usepackage{cite}
\usepackage{amsmath,amssymb,amsfonts}
\usepackage{graphicx}
\usepackage{textcomp}
\usepackage{xcolor}
\usepackage[euler]{textgreek}
\usepackage{color}
\usepackage{url}
\usepackage{todonotes}
\usepackage{multirow}
\usepackage{tabularx}
\usepackage{subfig}
\usepackage{enumitem}
\usepackage{authblk}

\begin{document}

\title{Cross Architectural Power Modelling}

\author{Kai Chen\textsuperscript{1},
Peter Kilpatrick\textsuperscript{1}, 
Dimitrios S. Nikolopoulos\textsuperscript{2}, and
Blesson Varghese\textsuperscript{1}
}
\affil{
\textsuperscript{1}\textit{Queen's University Belfast, UK};
\textsuperscript{2}\textit{Virginia Tech, USA}\\
E-mail: kchen10@qub.ac.uk; p.kilpatrick@qub.ac.uk; dsn@vt.edu; b.varghese@qub.ac.uk
}

\maketitle

\thispagestyle{plain}
\pagestyle{plain}

\begin{abstract}
Existing power modelling research focuses on the model rather than the process for developing models. An automated power modelling process that can be deployed on different processors for developing power models with high accuracy is developed. For this, (i) an automated hardware performance counter selection method that selects counters best correlated to power on both ARM and Intel processors, (ii) a noise filter based on clustering that can reduce the mean error in power models, and (iii) a two stage power model that surmounts challenges in using existing power models across multiple architectures are proposed and developed. The key results are: (i) the automated hardware performance counter selection method achieves comparable selection to the manual method reported in the literature, (ii) the noise filter reduces the mean error in power models by up to 55\%, and (iii) the two stage power model can predict dynamic power with less than 8\% error on both ARM and Intel processors, which is an improvement over classic models. 


\end{abstract}

\section{Introduction}
\label{sec:introduction}
Power monitoring has become a significant task for data-centre management since power consumption affects the cost of electricity and stability of server farms~\cite{ref1}. 
Although direct power measurement obtained by physical meters or model-based interfaces has been widely supported~\cite{lrpm-3}, it is not feasible for measuring power of individual hardware/software components. Fine-grained power measurement plays a significant role in runtime energy/performance management/optimisation~\cite{lrpm-2} 
 and energy-aware billing in data centers~\cite{utilisationlevelmodel-2}. For instance, both the model-based energy interface of the Intel Sandy Bridge server\footnote{\url{http://www.intel.com/content/www/us/en/processors/architectures-software-developer-manuals.html}} and the physical power meter of ARM Odroid-XU3 board\footnote{\url{http://www.hardkernel.com}} can measure the power of the entire processor rather than of individual computing cores or the executing programs.

A large proportion of power models rely on multiple hardware activities of the processor represented by hardware performance counters (or hardware counters) for estimating power~\cite{lrpm-1,lrpm-3}. 
However, the hardware counters necessary to build an accurate power model may substantially differ across processors due to the differences in the instruction set, pipeline, cache architecture and on-chip interconnect. The hardware counters are usually selected manually on the basis of experimental knowledge of the processor~\cite{ppep-1,lrpm-2}. 
 Typically, all possible hardware counters that can be obtained from a processor are extensively explored using a cumbersome trial and error approach after which a suitable few are selected~\cite{lrpm-arm-1}. Such an approach does not easily scale for various processor architectures since a different set of hardware counters will be required to model power for each processor.  

Currently, there is little research that develops automated methods for selecting hardware counters to capture processor power over multiple processor architectures. Automated methods are required for easily building power models for a collection of heterogeneous processors as seen in traditional data centers that host multiple generations of server processors, or in emerging distributed computing environments like fog/edge computing~\cite{offload-2} and mobile cloud computing (in these environments, an application may be distributed across different form factor processors, such as Intel Xeon processors~\cite{hcpm-xeon-1}, and low power processors, for example, ARM~\cite{lrpm-arm-1}). 
 Moreover, automated methods can be used to build power models for new processors with architectures that are currently not known.

This paper focuses on automating the power modelling process for different processor architectures. It is a power modelling process we propose that is cross architectural rather than any specific power model. 
The research contributions are:
\begin{itemize}[leftmargin=0.4cm]
\item The design and implementation of an automated hardware counter selection method to simplify the selection process without sacrificing the accuracy of the power model. Preliminary exploration of the hardware counter selection method has presented previously~\cite{bv-01}. 
\item The development of a clustering based noise filtering technique. The technique identifies and appropriately deals with noise from power related data obtained from multiple heterogeneous processors so as to improve the accuracy of power models that rely on the data.
\item The proposal of a novel power model, referred to as the Two Stage Power Model which takes advantage of both Linear Regression and Support Vector Machines.
\end{itemize}

The remainder of this paper is organised as follows. 
Section~\ref{sec:definitions} presents the notation and hardware platform employed. 
Section~\ref{sec:selectionmethod} proposes a method for selecting hardware counters. 
Section~\ref{sec:noisefilter} develops a technique for filtering noise. 
Section~\ref{sec:evaluationofmethodandtechnique} evaluates the selection method and filtering technique.
Section~\ref{sec:twostagepowermodel} proposes a novel two stage power model and is evaluated against classic power models. 
Section~\ref{sec:relatedwork} presents related research.
Section~\ref{sec:conclusions} concludes this paper.

\section{Definitions}
\label{sec:definitions}
\begin{table}[t]
	\caption{16 MPI and OpenMP benchmarks used}
	\label{table1}
\begin{center}
    \begin{tabular}{c p{4.7cm} p{1.5cm}}
		\hline	
		\textbf{Benchmark}	&	\textbf{Description}	& \textbf{MPI/OpenMP} \\
		\hline	
		\hline	
		BLS	&	Buffon-Laplace Simulation \cite{buffon-1} 	& 	MPI \\
		MCS	&	Monte-Carlo Simulation \cite{matvec-1}		&	MPI \\
		POI	&	Solve Poisson Equation in 2D 				& 	MPI\\
		RING	&	Ring Communication~\cite{ring-1}			& MPI\\
		WAV	&	Solve Wave Equation							& MPI\\
		SRCH	&	Searches integers in $[A,B]$ for a $J$ so that $F(J) = C$	& MPI\\
		\hline
		FFT	&	Fast Fourier Transform \cite{fft-1}			&	OpenMP	\\
		SGEFA	&	Solve Linear System $Ax = B$ \cite{sgefa-1}	& OpenMP\\
		ZIG	&	Obtains an exponentially distributed single precision real value~\cite{ziggurat-1}		& OpenMP\\
		MD	& 	Molecular Dynamics Simulation\cite{md-1}		& OpenMP\\
		\hline
		PRM	&	Generate Prime Numbers						& Both\\
		QUAD	&	Approximates an integral using a quadrature rule	& Both\\
		CSAT	&	Exhaustive search for solutions of the circuit satisfiability problem~\cite{satisfy-1}	& Both\\	
		\hline
	\end{tabular}
	\end{center}
\end{table}

In this section, we consider the mathematical notation and the hardware platform employed in this work.

\textbf{\textit{Notation:}} 
We define classic power models that are used for estimating dynamic power of processors and the concept of vectors and vector groups used in this paper. 

\subsubsection{Classic Power Models}
Consider a processor power model in which the estimated power, $P$, is the sum of the idle power of the processor (static), and the power required for various activities of the processor (dynamic). Thus,
\begin{equation}
P = P_{static} + P_{dynamic}
\end{equation}

In this paper, power modelling is explored in the context of dynamic power (a function of the volume of hardware activities on the processor). Hardware activity on processors, such as ARM and Intel, can usually be obtained from a catalogue of hardware performance counters (or hardware counters).
Consider $n$ hardware counters that can be obtained from a processor during the time interval $t_i$, denoted as $e_{i_1}$, $e_{i_2}$, $\cdots$ , $e_{i_n}$, and $P_{i_{dynamic}}$ is dynamic power. We consider the following three classic power models. 

\textit{a. Linear Regression Power Model (LRPM)}: In this model, dynamic power is defined as  
\begin{equation}
\label{equation:powerdynamic}
P_{dynamic} = \sum\limits_{i=1}^n c_{i}e_{i}
\end{equation}
where $c_i$ is the coefficient of the $i^{th}$ hardware counter.

\textit{b. Neural Network Power Model (NNPM)}: Compared to the LRPM which only captures linear relationships, NNPM can model both linear and non-linear relationships. A fully connected feed-forward NN using hardware counters is employed. 


\textit{c. Support Vector Machine Power Model (SVMPM)}:
This model captures both linear and non-linear relationships between dynamic power and the hardware counters using kernel tricks. A set of hyperplanes are fitted using the training data. Then these hyperplanes are used to estimate the dynamic power for given hardware counters. 

The ideal configuration of input parameters for both NNPM and SVMPM was chosen by extensively exploring the space. The configuration chosen for these models are those parameters that provide an accurate estimate of dynamic power.  

\subsubsection{Vectors}
We define a vector as
$V_{i}~=~\{P_{i_{dynamic}}, e_{i_1}, e_{i_2}, \cdots , e_{i_n}\}$, 
where the measured dynamic power during any given time interval corresponds to the set of hardware counter values obtained in the interval. 
Each vector is normalised to bring values of all variables in the same range between 0 and 1. The normalised vector of $V_{i}$ is represented as 
$\hat{V_{i}}~=~\{\hat{P}_{i_{dynamic}}, \hat{e}_{i_1}, \hat{e}_{i_2}, \cdots , \hat{e}_{i_n}\}$,
where 
$\hat{P}_{i_{dynamic}} =
P_{i_{dynamic}}$, $\hat{e}_{i_1} = \frac{e_{i_1}-min(e_1)}{max(e_1)-min(e_1)}$, $\cdots$ , $\hat{e}_{i_n} = \frac{e_{i_n}-min(e_n)}{max(e_n)-min(e_n)}$.

\subsubsection{Vector Groups}
\label{sssec:vectorgroups}
Normalised vectors are clustered into a set of Vector Groups (VGs). Clustering is performed such that each VG consists of similar vectors. Two vectors $\hat{V_{i}}$ and $\hat{V_{j}}$ are defined to be similar if the following conditions are satisfied:
\begin{equation}
\label{avbv-1}
a_V \leq \frac{\hat{P}_{i_{dynamic}}}{\hat{P}_{j_{dynamic}}} \leq b_V
\end{equation} 
and
\begin{equation}
\label{avbv-2}
a_V \leq \frac{\hat{e}_{i_k}}{\hat{e}_{j_k}} \leq b_V
\end{equation}
where $k=1, 2, \cdots n$, and $a_V$ and $b_V$ are user-defined bounds to determine similarity. For a given $a_V$, $b_V$ is as follows:
\begin{equation}
b_V~=~\frac{1}{a_V}
\end{equation}
The clustering algorithm will be presented in Section~\ref{sec:noisefilter}.

\textbf{\textit{Platform:}} 
Distributed computing environments such as those employed in Fog/Edge computing make use of both the cloud data center and edge nodes. Typically, data center servers, for example Amazon cloud servers, make use of Intel Xeon processors\footnote{\url{https://aws.amazon.com/ec2/instance-types/}}, which are designed for high-performance computing. On the other hand edge nodes make use of low power processors, such as ARM\footnote{\url{http://www.arm.com/products/iot-solutions/mbed-iot-device-platform}}. Next generation power models will need to work for emerging distributed computing environments and therefore, both an Intel Xeon processor representing servers used in data centers and an ARM processor representing smaller form factor Edge nodes are used. 

The first processor is the Intel Xeon Sandy Bridge server comprising two Intel Xeon E5-2650 processors with 8 cores on each processor, 32KB/32KB  I/D-Cache  per  core,  2MB  shared L2 cache per 8 cores and 20MB shared L3 cache per package and running CentOS 6.5. We measure power consumption of the power lane which supports the multi-core processor and the on-chip caches (L1/L2/L3) by directly reading the on-chip energy counter through Intel's RAPL interface.

The second processor is the ODROID-XU+E\footnote{\url{http://www.hardkernel.com}} board which has one ARM Big.LITTLE architecture Exynos 5 Octa processor. There are four Cortex-A15 cores and four Cortex-A7 cores, 32KB/32KB I/D-Cache per core, NEONv2 floating point support per core, VFPv4 support per core, one PowerVR SGX 544 MP3 GPU, and 2 GBytes of LPDDR3 DRAM. A 2 MByte L2 cache is shared between all Cortex-A15 cores and a 512 KByte L2 cache is shared between all Cortex-A7 cores. The ODROID board provides power meters to measure the power of different components, including the Cortex-A7 and Cortex-A15 cores. We use the power meter, which measures the power of the Cortex-A15 cores, including their L1 caches and shared L2 cache. The system runs Ubuntu 14.04 LTS. 

The hardware counters are obtained from a real-time profiling framework, namely Performance API (PAPI)\footnote{\url{http://icl.cs.utk.edu/papi/}} \cite{PAPI-1}. Power is obtained from the on-chip power sensor on ARM and from the RAPL interface on Intel. We  employ 16 scientific benchmarks (source code was available from a public repository\footnote{\url{http://people.sc.fsu.edu/~jburkardt/cpp\_src/cpp\_src.html}}) as shown in Table \ref{table1}. On ARM, we used the Cortex-A15 cores at their maximum frequency of 2.0GHz to execute the benchmarks. The Cortex-A7 cores at their maximum frequency of 1.4 GHz are used to obtain vectors (which include measured power and hardware counters). Similarly, on Intel, we used one processor at its maximum frequency of 2.0GHz to execute the benchmarks and the second processor at its maximum frequency of 2.0GHz to obtain vectors.  

The vectors are obtained using a Start-Stop (SS) sampling method, which is widely reported in the literature~\cite{powerproxies-1,powercontainers-1,ppep-1}. 
In this method, vectors are continuously sampled during the execution of benchmarks approximately every 1 second. The hardware counters of each vector are calculated as 
$e_i = \frac{e_{e_i}-e_{b_i}}{t}$
, where $e_{b_i}$ and $e_{e_i}$ are values of the $i^{th}$ hardware counter obtained at the beginning and the end of each sampling interval, respectively. $t$ is the length of the sampling interval and it is approximately 1 second.

Power is measured from the ARM and Intel processors differently (hardware counters are obtained using PAPI on both platforms). On ARM, power consumed in the sampling interval is calculated as
$P = \frac{P_b+P_e}{2}$
, where $P_b$ and $P_e$ are values of power read from the on-chip power sensor at the beginning and the end of each sampling interval, respectively.
On Intel, power consumed in the sampling interval is obtained as 
$P = \frac{E_e-E_b}{t}$
, where $E_b$ and $E_e$ are values of energy read from the RAPL interface at the beginning and the end of each sampling interval, respectively. Unlike the ARM platform, only energy values can be obtained on the Intel platform.

\section{Hardware Counter Selection (HCS) Method}
\label{sec:selectionmethod}

A generic and automated Hardware Counter Selection (HCS) method that can be employed on multiple processors is presented. The method selects a set of six hardware counters (a maximum of only six hardware counters can be obtained simultaneously using PAPI on the ARM and Intel processors used in this research; additional hardware counters can be obtained using multiplexing, but introduces overhead and is not used) from those available that best correlates to power for a given processor. The method utilises a Random Forest (RF) algorithm that maps the hardware counters to power. RF is chosen due to its accuracy in regression~\cite{RF-1}. While an RF based power model will not be feasible for on-line power monitoring due to its high computing complexity, it quantifies the relative importance of each hardware counter to power during the model fitting process. Hence, we leverage this characteristic of RF algorithms to build the automatic hardware counter selection method that works off-line. The selection method may need to be executed only once (or a limited number of times) for a processor to determine which hardware counters best correlate to power.

\begin{algorithm}[t]\small
\caption{Hardware Counter Selection (HCS) Method}
\label{algo:selection}
\begin{algorithmic}[1]
\Procedure{select\_counters}{$all\_vectors$, $n$, $ntree$}
\State $counters\_selected \gets list\left(\right)$
\For{$i=1$ to $M$} \Comment{The entire dataset is partitioned into $M$ subsets }
   
   \State$part\_vectors \gets extract\left(all\_vectors, i, M\right)$ \Comment{Extract the $i^{th}$ subsets from overall M subsets}
   \State$rfes \gets $ randomForest$ \left(part\_vectors, ntree\right)$ 
   \State $counters\_importance[i] \gets rfes.importance$ 
   
\EndFor
\State Find n hardware counters with largest average value of importance. 

\State \textbf{Return} n events with largest average importance value 
\EndProcedure
\end{algorithmic}
\end{algorithm}

The HCS method is designed to generate a list of hardware counters that are most relevant to power estimation as shown in Algorithm~\ref{algo:selection}. The key design principle is that the HCS method should be suitable for all applications and the hardware counters selected by the approach should be independent of applications. The dataset is partitioned and hardware counters for each subset are obtained to minimise dependence on the dataset.
The inputs to the HCS method are:

1) $all\_vectors$ is the set of all vectors (and all hardware counters available on a processor) obtained from executing the benchmarks. The PAPI multiplexing function\footnote{Multiplexing has large overheads and is therefore only employed in the HCS method and not for building the power model} is used.

2) $n$ is the number of hardware counters to be selected. In our case, we use six, which is the maximum number of counters obtained simultaneously from PAPI.

3) $ntree$ is the number of trees that are used to build the random forest model. This parameter is empirically determined.

The algorithm firstly partitions $all\_vectors$ into a set of subsets (i.e. $M$ subsets) (Lines 3-4). During each iteration of the $for$ loop (line3) for each subset $i$ that is extracted from the overall $M$ subsets (line 4), a Random Forest model is used to map hardware counters to power (Line 5). The importance of each hardware counter for a given partition is obtained and stored in the $counters\_importance$ (Line 6). Finally, $n$ hardware counters with largest average importance values are found (Line 7) and returned (Line 8).

If $all\_vectors$ are used without partitioning, then the selected counters will depend on the entire dataset, which is not ideal for a general HCS method. 
We generate a set of subsets by partitioning $all\_vectors$. Then the importance of hardware counters for each subset is obtained. We select hardware counters that have the largest average importance values. Such a method results in a dataset independent method.

\section{Noise Filtering (NF) Technique}
\label{sec:noisefilter}
A second problem when designing a hardware counter-based power model is related to \textit{filtering noisy data} since it affects the accuracy of estimation. Noisy data in power modelling can be an artefact of measurement employed for obtaining (sampling) vectors (Section~\ref{sec:definitions}) or due to the instability of physical sensors/interfaces. Hence, we design and develop a noise filter that is suitable for multiple processors. A method to identify noisy vectors and an appropriate mechanism to deal with noise are considered in the proposed filter. 

\begin{figure}[t]
\centering
	\subfloat[process becomes idle]
	{	\label{noise1a}
		\includegraphics[width=0.23\textwidth]
		{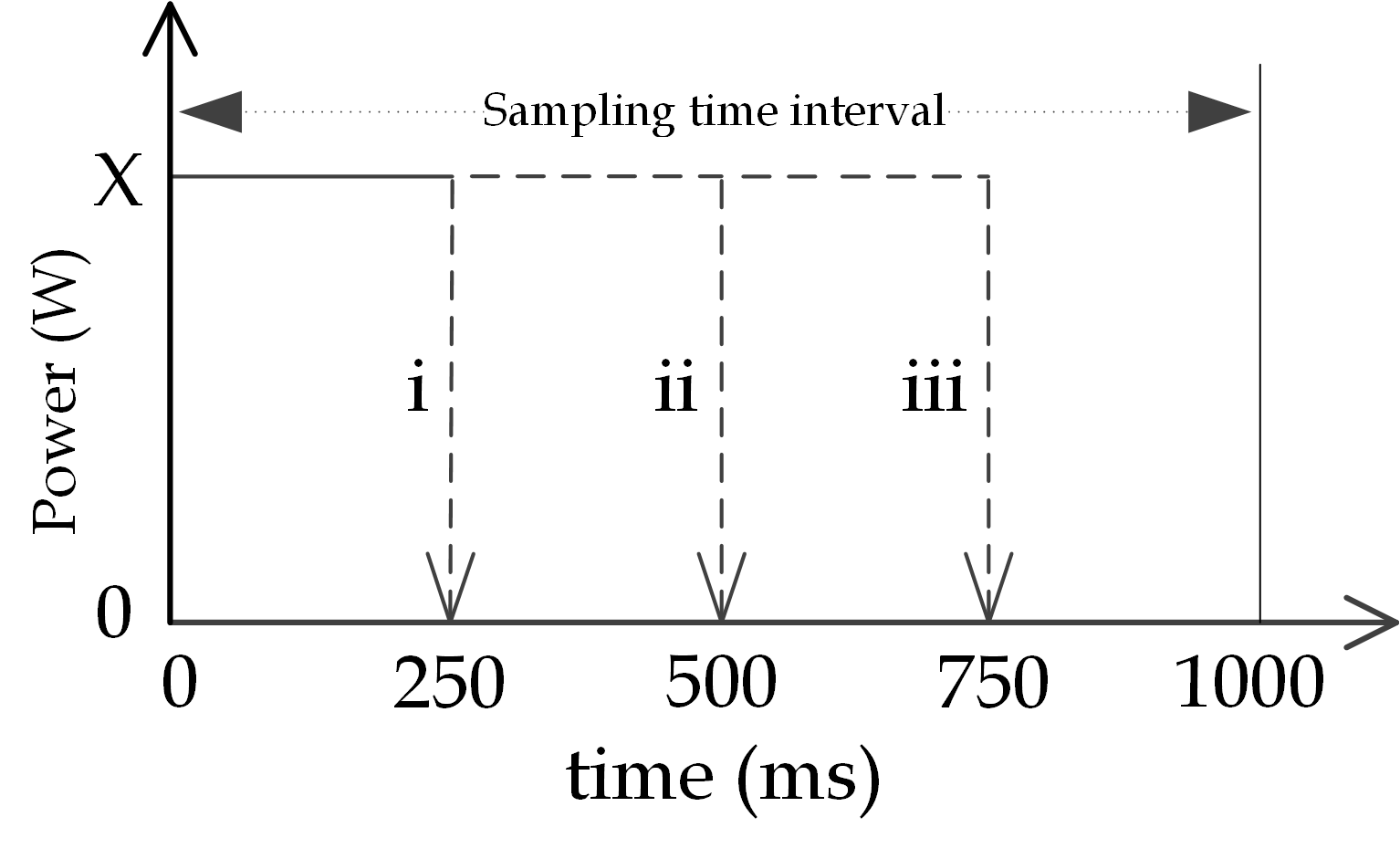}
	}
	\hfill
	\subfloat[process changes state]
	{	\label{noise1b}
		\includegraphics[width=0.23\textwidth]
		{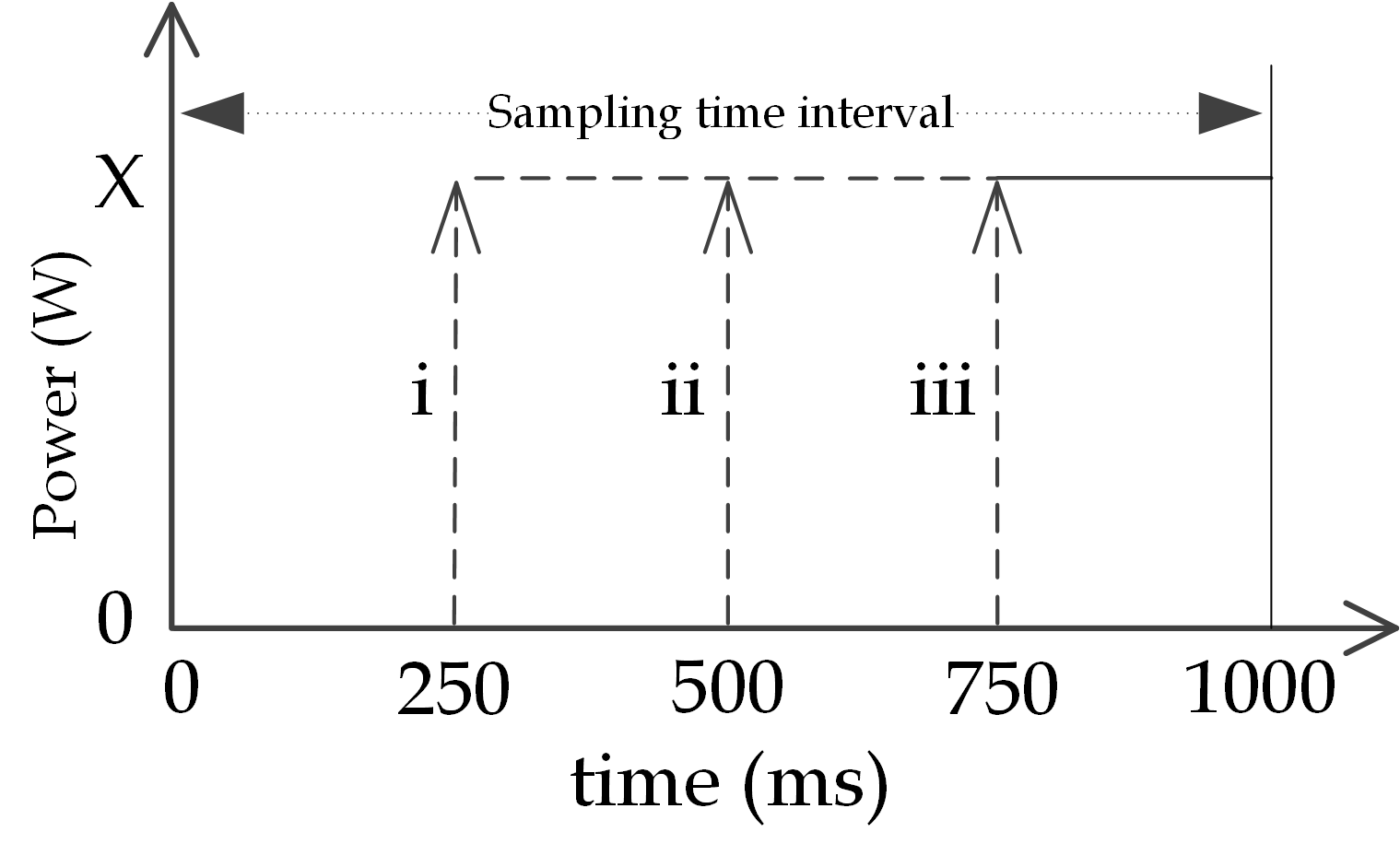}
	}
	\caption{Type I noise vectors during the sampling time interval}
	\label{fig:noise_1}
\end{figure}   

\begin{figure}[t]
\centering
	\subfloat[power of process decreases]
	{	\label{noise2a}
		\includegraphics[width=0.23\textwidth]
		{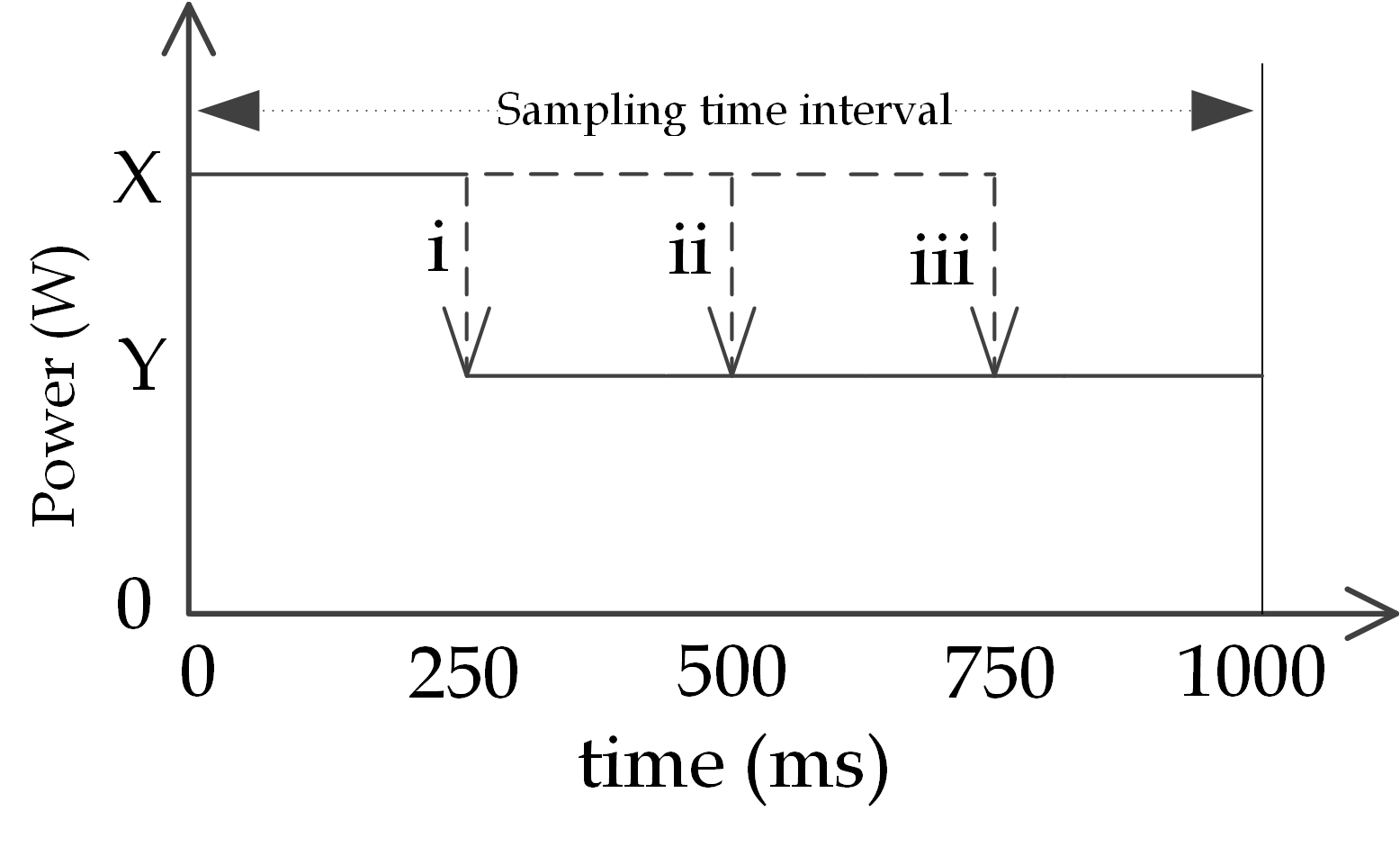}
	}
	\hfill
	\subfloat[power of process increases]
	{	\label{noise2b}
		\includegraphics[width=0.23\textwidth]
		{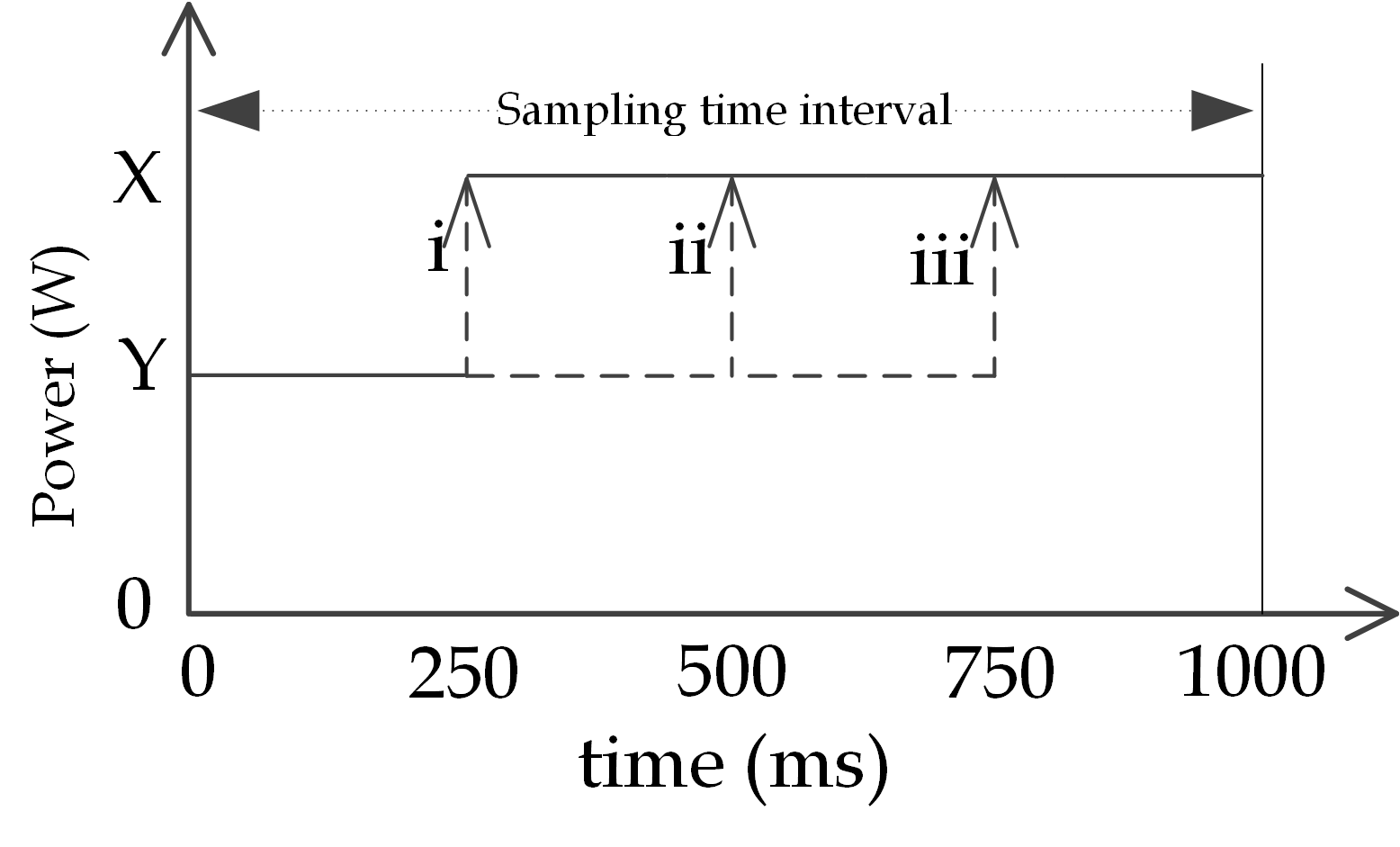}
	}
	\caption{Type II noise vectors during the sampling time interval}
	\label{fig:noise_2}
\end{figure}   

\subsection{Types of Noise}
On both the ARM and Intel processors the Start-Stop (SS) sampling method for a given time interval (1 second) considered in Section~\ref{sec:definitions} is used. However, on the ARM processor we read power values from the sensors, but on Intel processor we obtain energy values through the RAPL interface. Reading power values on ARM poses two problems resulting in noisy data (refer Figure~\ref{fig:noise_1} and Figure~\ref{fig:noise_2}). 

The first problem is highlighted in Figure~\ref{fig:noise_1} when a process completes execution or an idle process resumes execution in the sampling interval. The sampling interval for obtaining power is 1~second and the power sensor is read at the beginning and end of the interval. Consider that at the beginning of the time interval the power sensor provided a reading of $X$~Watts ($X$~W) and at the end of the time interval provided $0$~W or vice versa. Then the SS sampling method would record the power value for this time interval as $X/2$~W (mean of $0$~W and $X$~W). However, it is possible that power has changed during this time interval such that the mean power is not $X/2$~W (this change is illustrated as i, ii or iii when power values change at 250~ms, 500~ms or 750~ms respectively in the figure) for the time interval. Conventional SS method cannot capture this change and results in power values that do not correspond to the hardware counters. We refer to this as \textit{Type I} noisy data.

The second problem, highlighted in Figure~\ref{fig:noise_2}, is when the power consumption behaviour of a process changes such that it consumes less or more power in the sampling interval. Consider that at the beginning of the time interval the power sensor provided $X$~W and at the end of the interval provided $Y$W or vice versa. Then the SS sampling method records power for this time interval as $(X+Y)/2$~W.
However, power consumed by the process could have changed during this time interval as illustrated by i, ii or iii when power values change at 250~ms, 500~ms or 750~ms, respectively, in the figure. Again the conventional SS method cannot capture this change and results in power values that do not correspond to the hardware counters. We refer to this as \textit{Type II} noisy data. 

Both Type I and Type II noise are caused when the power consumption behavior of a program changes during the sampling interval. Type I can be considered as a specific case of Type II. However, how they are identified and dealt with in the proposed technique are different as presented in Section~\ref{sec:noisefilter}. The power referred to in Figure~\ref{fig:noise_1} is dynamic power and may be zero when the processor is idle. 

Due to the instability of power sensors and associated interfaces, a small set of vectors may have measured power values significantly different from their actual power consumption. Hardware counters that correspond to the utilisation of processors are widely accepted for predicting dynamic power of a processor. However, if two vectors have similar hardware counter values but significantly different measured power values, at least one of the power values has to be treated as an anomaly. We refer to this as \textit{Type III} noisy data. 

Type I and Type II noise are inherent to methods that measure power (not energy). Hence, they only appear in the dataset profiled from the ARM processor where an on-chip power sensor is used rather than Intel on which energy is measured during a sampling interval using RAPL. Type III noise, which may be due to instability of power sensors, is found in datasets from both ARM and Intel.

\subsection{Noise Filter}
The filter we propose targets the above three types of noise by identifying them and appropriately dealing with them. Our technique for addressing Type I and Type II noise is by modifying the power value in a vector, which is considered in this section. For Type III noise we simply remove the anomalous vectors from the data. The Noise Filter (NF) technique comprises the following six steps:
\subsubsection*{Step 1 - Clustering vectors} 
In the first step, all vectors are clustered into Vector Groups according to power and hardware counters on the basis of Equation~\ref{avbv-1} and Equation~\ref{avbv-2}~(described in Section~\ref{sec:definitions}) using Algorithm~\ref{algo:clustering}. For each vector, $\hat{V}_i$, if there is a VG such that all vectors in this VG are similar to $\hat{V}_i$, then $\hat{V}_i$ will be added to the VG (Lines 4-5). Otherwise, a new VG is created and $\hat{V}_i$ is added into the new VG (Line 6-8). 

\begin{algorithm}[t] \small
\caption{Clustering algorithm}
\label{algo:clustering}
\begin{algorithmic}[1]
\Procedure{clustering\_algorithm}{$all\_vectors$}
\State $n \gets sizeof\left(all\_vectors\right)$
\For{$i=0$ to $n$}
   \If{$\exists VG_p$ whose vectors are all similar with $all\_vectors[i]$}
       \State Add $all\_vectors[i]$ to $VG_p$
   \Else
      \State Create a new VG: $VG_{q}$
      \State Add $all\_vectors[i]$ to $VG_{q}$
   \EndIf   
\EndFor
\EndProcedure
\end{algorithmic}
\end{algorithm}

\subsubsection*{Step 2 - Identifying normal vectors} 
Vectors from the sampling data that are not noisy are referred to as \textit{normal vectors}. For a given sample of data if consecutive vectors are clustered in to the same VG, then there is less probability that these vectors contain noisy power values. Therefore, they are considered to be normal vectors. Additionally, all vectors in this VG are considered to be normal. 

\subsubsection*{Step 3 - Identifying and modifying \textit{Type I} noise vectors} 
\begin{figure}
\centering
\includegraphics[width=0.45\textwidth]
{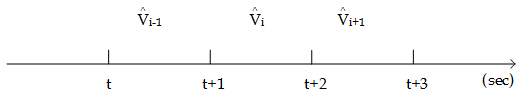}
\caption{Neighbouring vectors, $\hat{V}_{i-1}$ and $\hat{V}_{i+1}$ of $\hat{V}_i$}
\label{fig:neighbourvectors}
\end{figure}

Consider three consecutive vectors, $\hat{V}_{i-1}$, $\hat{V}_i$ and $\hat{V}_{i+1}$ as shown in Figure~\ref{fig:neighbourvectors}. We define $\hat{V}_{i-1}$ and $\hat{V}_{i+1}$ as \textit{neighbour vectors} of $\hat{V}_i$, denoted as $\hat{V}_{nr}$. $\hat{V}_i$ is identified as a Type I noise vector, if the following three conditions are met:

\noindent i. if at least one of the neighbour vectors is a normal vector, 

\noindent ii. the dynamic power $\hat{P}_{i_{dynamic}} \approx \frac{\hat{P}_{{nr}_{dynamic}}}{2} $

\noindent iii. the ratios of the hardware counter of $\hat{V}_i$ and the normal neighbour vector are similar, but are not equal (approx.) to $0.5$ 
\begin{equation*}
\frac{\hat{e}_{i_1}}{\hat{e}_{{nr}_1}} \approx \frac{\hat{e}_{i_2}}{\hat{e}_{{nr}_2}} \approx \cdots \approx \frac{\hat{e}_{i_n}}{\hat{e}_{{nr}_n}} \neq 0.5
\end{equation*}

If $\hat{V}_i$ is a Type I noise vector, then it is modified as 
\begin{equation}
\label{form:type1}
\hat{P}_{i_{dynamic}} = \hat{P}_{nr_{dynamic}} \times \frac{\hat{e}_{i_1}}{\hat{e}_{nr_1}}
\end{equation}
The $\frac{\hat{e}_{i_1}}{\hat{e}_{nr_1}}$ ratio corrects the measured power of $V_i$ and makes it a normal vector. 

\subsubsection*{Step 4 - Identifying and modifying \textit{Type II} noise vectors}
Consider three consecutive vectors, $\hat{V}_{i-1}$, $\hat{V}_i$ and $\hat{V}_{i+1}$ as shown in Figure \ref{fig:neighbourvectors}. $\hat{V}_i$ is identified as a Type II noise vector, if the following three conditions are met:

\noindent i. both $\hat{V}_{i-1}$ and $\hat{V}_{i+1}$ are normal vectors,

\noindent ii. the dynamic power $\hat{P}_{i_{dynamic}} \approx \frac{\hat{P}_{{i-1}_{dynamic}} + \hat{P}_{{i+1}_{dynamic}}}{2} $

\noindent iii. the hardware counters $\frac{\hat{e}_{i_1}-\hat{e}_{{i+1}_1}}{\hat{e}_{{i-1}_1} - \hat{e}_{{i+1}_1}} \approx \frac{\hat{e}_{i_2}-\hat{e}_{{i+1}_2}}{\hat{e}_{{i-1}_2} - \hat{e}_{{i+1}_2}} \approx \cdots \frac{\hat{e}_{i_n}-\hat{e}_{{i+1}_n}}{\hat{e}_{{i-1}_n} - \hat{e}_{{i+1}_n}} \neq 0.5$

If $\hat{V}_i$ is Type II noise vector, then it is modified as 
\begin{equation}
\label{form:type2}
\hat{P}_{i_{dynamic}} = r \times \hat{P}_{{i-1}_{dynamic}} + (1-r) \times \hat{P}_{{i+1}_{dynamic}} 
\end{equation}  
where $r = \frac{\hat{e}_{i_1} - \hat{e}_{{i+1}_1}}{\hat{e}_{{i-1}_1} - \hat{e}_{{i+1}_1}}$, $\hat{P}_{{i-1}_{dynamic}}$ and $\hat{P}_{{i+1}_{dynamic}}$ are dynamic power of the neighbour vectors of $\hat{V}_i$, $\hat{e}_{{i-1}_1}$ and $\hat{e}_{{i+1}_1}$ are normalised values of 
$\hat{e}_1$ of the neighbour vectors.

Type I noise occurs when the behavior of a program changes from execution to idle or from idle to execution. The dynamic power consumption changes from one value ($X$) to 0 or vice versa. To identify this noise, two vectors, namely a given vector and one of its neighbouring vectors need to be considered. Type II noise occurs when a program changes from one phase to another. The dynamic power consumption changes from one value ($X$) to a different value ($Y$) or vice versa. To identify this noise, three vectors, namely a given vector and two of its neighbouring vectors need to be considered. An additional vector is required to identify Type II because, unlike Type I, one of the power values is not zero.

\subsubsection*{Step 5 - Re-clustering vectors}
In Step 1, we used both hardware counters and dynamic power for clustering. However, in this step, we only use hardware counters for clustering with the intention of identifying vectors with anomalies in power values. Algorithm~\ref{algo:clustering} can be employed by simply excluding measured dynamic power from the input vectors.  
\subsubsection*{Step 6 - Identifying and Removing \textit{Type III} noise vector}
A vector $\hat{V}_i$ is identified as a Type III noise vector if its power value is significantly different from the power of normal vectors identified in Step 2 that are grouped into the same VG as $\hat{V}_i$ by the clustering process in Step 5. Type III noise vectors are simply removed from the data. 


\section{Evaluating the HCS Method and\\the NF Technique}
\label{sec:evaluationofmethodandtechnique}
In this section, we evaluate the hardware counter selection method of Section~\ref{sec:selectionmethod} and the noise filtering technique  of Section~\ref{sec:noisefilter}. 
The HCS method is evaluated by comparing the power estimation accuracy of classic power models, such as Linear Regression Power Models (LRPMs) presented in Section~\ref{sec:definitions}, when using hardware counters obtained from our selection method against a baseline using hardware counters reported in the literature. The NF technique is evaluated by comparing the prediction error of the LRPM with and without using the noise filter we propose. LRPMs are chosen for this evaluation since they are popularly used~\cite{lrpm-1,lrpm-2,lrpm-3}.

For evaluating both the HCS method and the NF technique we use a rigorous training and testing strategy on the LRPM. All vectors obtained from profiling the execution of the benchmarks are equally partitioned into four parts. Then we use a combination of three parts to train the LRPM. The trained model is used to test: (i) vectors from the three parts used to train the model (75\% of the vectors). We refer to these vectors as \textit{`\textbf{Known}'} vectors because they are known to the model through the training process, and (ii) vectors from the fourth part which were not used for training (25\% of the vectors). We refer to these vectors as \textit{`\textbf{Unknown}'} vectors.
A preprocessing step was included, such that no unknown vectors are similar to the known vectors. The training and testing strategy is repeated four times for different combinations of partitioned vectors. We note that the evaluation based on unknown vectors presents the accuracy of power models when they are used to estimate power of vectors that are not the same or similar in the training data. Given that a large number of applications are emerging on modern processors it is impossible to obtain a training dataset that contains representative vectors from all applications. Therefore, an ideal power model must be designed to work well for unknown vectors.

We evaluate both the HCS method and NF technique using Linear Regression (LRPM), Support Vector Machine (SVMPM) and Neural Network (NNPM) power models presented in Section~\ref{sec:definitions}. However, since the results obtained are similar we present results that are based only on LRPM.

\subsection{Evaluating the HCS method}
The HCS method is evaluated by investigating the quality of the hardware counters selected by the HCS method. It was empirically found that the HCS method is not sensitive to $ntree$ (this aligns with findings of previous research~\cite{RF-3,RF-4}). Furthermore, considering the computation overhead due to employing larger values of $ntree$, we use $ntree=2$ and $ntree=16$ on ARM and Intel, respectively (the values of $ntree$ are empirically obtained).

\begin{table}[t]
	\caption{Hardware counters from the baseline and selected by the HCS method for different $ntree$ values on ARM}
	\label{tab:hardwarecountersarmntree}
\begin{center}   
	\begin{tabular}{c c c }
        \hline
        \multicolumn{3}{c}{\textbf{Hardware Counters}} \\ 
        \cline{1-3}
        \textbf{Baseline} &  \textbf{HCS ($ntree$=1)} & \textbf{HCS ($ntree$=2-512)}\\
		\hline	
		\hline	
		\texttt{PAPI\_TOT\_CYC}	&  \texttt{PAPI\_TOT\_CYC}	& \texttt{PAPI\_TOT\_CYC}  \\ 
		\texttt{PAPI\_TOT\_INS}	&  \texttt{PAPI\_TOT\_INS}	& \texttt{PAPI\_TOT\_INS} \\ 
		\texttt{PAPI\_L1\_DCA}	&  \texttt{PAPI\_L1\_DCA} & \texttt{PAPI\_L1\_DCA}	\\ 
		\texttt{PAPI\_L1\_ICA}	&  \texttt{PAPI\_L1\_ICA} & \texttt{PAPI\_L1\_ICA}	\\ 
        \texttt{PAPI\_L2\_TCM} &  \texttt{PAPI\_L2\_TCM} & \texttt{PAPI\_L2\_TCM}  \\ 
        \texttt{PAPI\_TLB\_IM}	&  - & -  \\ 
        -	& \texttt{PAPI\_L2\_DCA} & - \\ 
        - & - & \texttt{PAPI\_L1\_ICM}  \\ 
        
		\hline
	\end{tabular}
	\end{center}
\end{table}

\begin{table*}[t]
	\caption{Hardware counters from the baseline and selected by the HCS method on Intel}
	\label{tab:hardwarecountersintelntree}
\begin{center}   
	\begin{tabular}{c c c c c c }
        \hline
        \multicolumn{6}{c}{\textbf{Hardware Counters}} \\ 
        \cline{1-6}
        \textbf{Baseline} &  \textbf{HCS (ntree=1)} &  \textbf{HCS (ntree=2)} &  \textbf{HCS (ntree=4)} &  \textbf{HCS (ntree=8)} &  \textbf{HCS (ntree=16-512)} \\
		\hline	
		\hline	
		\texttt{PAPI\_TOT\_CYC}	&	\texttt{PAPI\_TOT\_CYC}	& \texttt{PAPI\_TOT\_CYC}	& \texttt{PAPI\_TOT\_CYC}	& \texttt{PAPI\_TOT\_CYC}	& \texttt{PAPI\_TOT\_CYC}	\\ 
		\texttt{PAPI\_TOT\_INS}	&	\texttt{PAPI\_TOT\_INS}	& \texttt{PAPI\_TOT\_INS}	& \texttt{PAPI\_TOT\_INS}	& \texttt{PAPI\_TOT\_INS}	& \texttt{PAPI\_TOT\_INS} \\ 
		\texttt{PAPI\_LD\_INS}	& - &	-	& - & \texttt{PAPI\_LD\_INS} & \texttt{PAPI\_LD\_INS} \\ 
        \texttt{PAPI\_SR\_INS}	&	\texttt{PAPI\_SR\_INS}	& - & - & - & \texttt{PAPI\_SR\_INS} \\ 
		\texttt{PAPI\_FP\_OPS}	& - &	- & - & - & -  \\ 
		\texttt{PAPI\_L3\_TCA}	&	- & - & - &	- & - \\ 
-	&	\texttt{PAPI\_REF\_CYC} & \texttt{PAPI\_REF\_CYC} & \texttt{PAPI\_REF\_CYC} & \texttt{PAPI\_REF\_CYC} & \texttt{PAPI\_REF\_CYC} \\ 
        - & \texttt{PAPI\_L3\_TCM} & \texttt{PAPI\_L3\_TCM}  & \texttt{PAPI\_L3\_TCM} & \texttt{PAPI\_L3\_TCM} & \texttt{PAPI\_L3\_TCM} \\ 
         - & - & - & \texttt{PAPI\_BR\_TKN} & - & - \\ 
        - & \texttt{PAPI\_BR\_NTK} & - & - & - & - \\ 
         - & - & \texttt{PAPI\_L3\_DCR} & \texttt{PAPI\_L3\_DCR} & \texttt{PAPI\_L3\_DCR} & - \\ 
        - & - & \texttt{PAPI\_BR\_UCN} & - & - & - \\ 
        
		\hline
	\end{tabular}
	\end{center}
\end{table*}

\begin{table}[ht]
	\caption{Hardware counters and descriptions}
	\label{tab:hardwarecounterdescription}
\begin{center}   
	\begin{tabular}{l l }
        \hline
        \textbf{Hardware Counters} & \textbf{Description}\\ 
		\hline	
		\texttt{PAPI\_BR\_CN}	& Conditional branch instructions \\ 
		\texttt{PAPI\_BR\_MSP}	& Conditional branch instructions mispredicted \\  
		\texttt{PAPI\_BR\_NTK}	& Conditional branch instructions not taken \\ 
		\texttt{PAPI\_BR\_TKN}	& Conditional branch instruction taken \\   
        \texttt{PAPI\_BR\_UCN} & Unconditional branch instructions \\ 
        \texttt{PAPI\_FP\_INS}	& Floating point instructions \\  
        \texttt{PAPI\_FP\_OPS}	& Floating point operations \\ 
        \texttt{PAPI\_L1\_DCA} & Level 1 data cache accesses \\ 
        \texttt{PAPI\_L1\_DCM} & Level 1 data cache misses \\ 
        \texttt{PAPI\_L1\_ICA} &  Level 1 instruction cache accesses \\
        \texttt{PAPI\_L1\_ICM} & Level 1 instruction cache misses \\  
        \texttt{PAPI\_L2\_DCA} & Level 2 data cache accesses \\ 
        \texttt{PAPI\_L2\_DCH} & Level 2 data cache hits \\ 
        \texttt{PAPI\_L2\_DCM} & Level 2 data cache misses\\
        \texttt{PAPI\_L2\_TCM}	& Level 2 cache misses\\
        \texttt{PAPI\_L3\_DCA} &  Level 3 data cache accesses\\
        \texttt{PAPI\_L3\_DCR} & Level 3 data cache reads\\
        \texttt{PAPI\_L3\_TCA} & Level 3 cache accesses\\
        \texttt{PAPI\_L3\_TCM} & Level 3 cache misses \\
        \texttt{PAPI\_LD\_INS} & Load instructions\\
        \texttt{PAPI\_REF\_CYC} & Reference clock cycles\\
        \texttt{PAPI\_SP\_OPS} & Optimized floating point operations\\
        \texttt{PAPI\_SR\_INS} & Store instructions\\
        \texttt{PAPI\_TLB\_DM} & Data translation lookaside buffer misses\\
        \texttt{PAPI\_TLB\_IM} & Instruction translation lookaside buffer misses\\
        \texttt{PAPI\_TOT\_CYC} & Total cycles\\
        \texttt{PAPI\_TOT\_INS} & Instructions completed\\
		\hline
	\end{tabular}
	\end{center}
\end{table}

Table~\ref{tab:hardwarecountersarmntree} and Table~\ref{tab:hardwarecountersintelntree} present the hardware counters (a description of all hardware counters is shown in Table~\ref{tab:hardwarecounterdescription}) 
that we use as a baseline and those selected by the HCS method for ARM and Intel, respectively.
The baseline is determined by reviewing existing research~\cite{lrpm-arm-1, lrpm-arm-2, lrpm-4, powercontainers-1} and by considering the characteristics of our experimental platform and hardware counter profiling tool PAPI. Although the processors employed are different, the general characteristics of the application and the hardware counters are the same. 
We selected six hardware counters because of the limit on the maximum number of counters that can be simultaneously obtained.

On the ARM and Intel processors we note that the hardware counters obtained from the HCS method are quite similar to those from the baseline (on ARM only one hardware counter is different and on Intel only two hardware counters differ). We infer from this that given different hardware processors our selection method can automatically obtain appropriate hardware counters that capture dynamic power. It is also observed that the hardware counters for the ARM and Intel processors are different (4 out of the 6 hardware counters differ). The HCS method we propose selects processor dependent hardware counters suitable for developing power models. 

To further evaluate the HCS method, we compare the accuracy of LRPMs using the hardware counters selected by  HCS  and those proposed by the literature, respectively. 

\begin{figure*}[ht]
\centering
    \subfloat[On ARM ($ntree = 2$)]
	{	\label{unknownvectorarm}
		\includegraphics[width=0.475\textwidth]
		{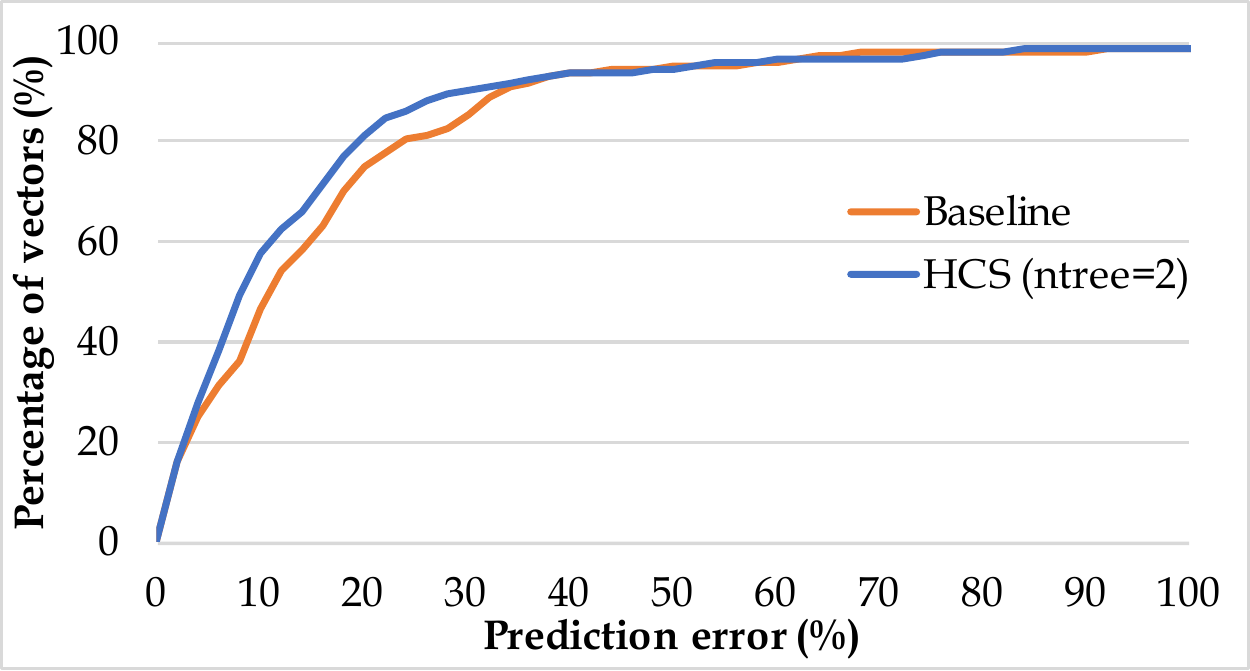}
	}	\hfill
	\subfloat[On Intel ($ntree$ = 16)]
	{	\label{unknownvectorintel}
		\includegraphics[width=0.475\textwidth]
		{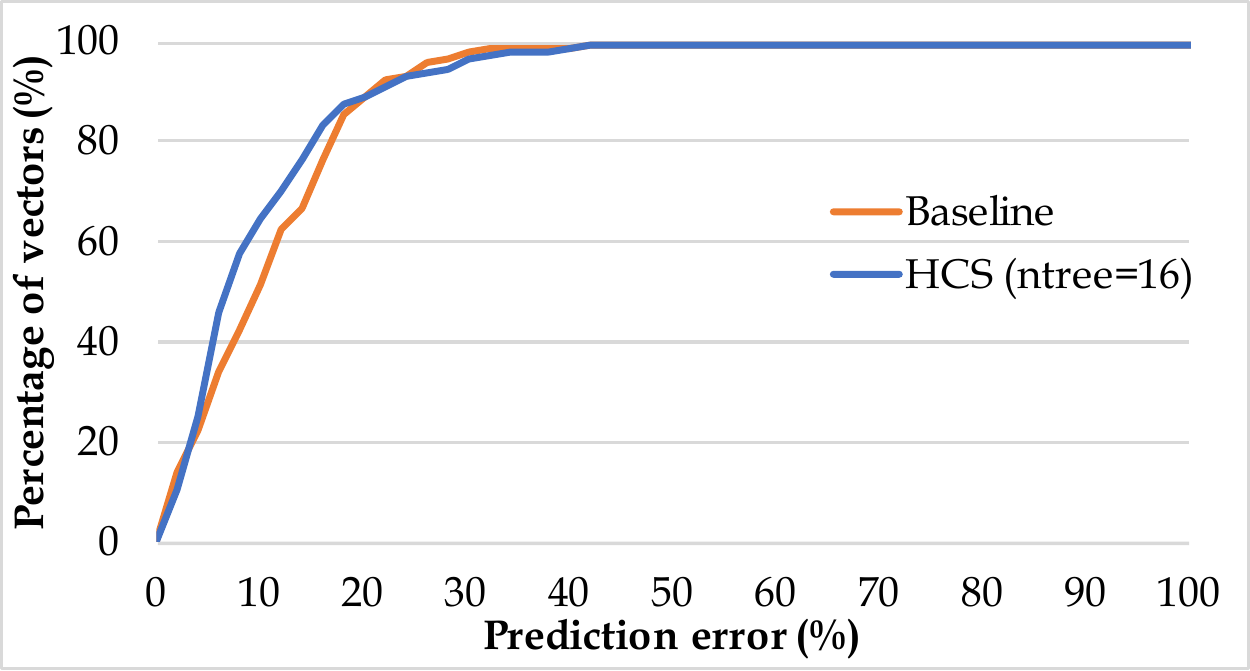}
	}
	\caption{Accuracy of LRPM for unknown vectors using hardware counters from the literature (baseline) and the HCS method}
	\label{fig:hardwarecounterselector}
\end{figure*}

Figure~\ref{fig:hardwarecounterselector} shows the cumulative distribution of $Error$ for LRPMs when used to estimate the power of Unknown vectors. The $x$ axis shows the $Error$ and the $y$ axis shows the percentage of vectors with error less than each value on $x$. In the best case, the HCS method on both processors performs better for Unknown vectors than the baseline. For example, for the ARM processor the HCS method based LRPM can accurately estimate the power of 58.1\% of Unknown vectors with $Error$ no more than 10\% which is nearly a 12\% improvement compared to the baseline based LRPM. The HCS method even in the worst case provides nearly similar accuracy to the baseline.

In summary, the automated method for simplifying the selection of hardware counters does not sacrifice the accuracy of power models. Existing research employs a manual and exhaustive exploration technique of all hardware counters. The proposed approach minimises human intervention and obtains similar accuracy when compared to the hand tweaked baseline. 

\subsection{Evaluating the NF Technique}
The NF technique is evaluated by considering the quality of filtering and the effect on the user-defined bounds. 
To evaluate the NF technique we first tested the LRPM using the hardware counters selected by the HCS method without using a noise filter. Then we tested the same model by using vectors that were filtered by our NF technique. This was performed on both ARM and Intel processors for both known and unknown vectors. Table~\ref{tab:noisefilter} highlights the key results captured as mean error and standard deviation from this evaluation. It is inferred that on ARM the mean error percentage when using the NF filter is reduced by more than half. On Intel the error percentage is reduced by nearly a third when using the NF filter. Thus the model is more stable in its prediction given that the standard deviation significantly reduces in both cases. The key result is that the NF technique reduces overall estimation error. 
It was empirically observed that the technique is not sensitive to the user-defined bounds (Section~\ref{sec:definitions}).


\begin{table}[ht]
	\caption{Mean error and standard deviation of vectors with and without using the NF technique}
	\label{tab:noisefilter}
\begin{center}
\centering
    
   \resizebox{0.48\textwidth}{!}{%
	\begin{tabular}{l c c}
		\hline	
				&	\textbf{Mean Error (\%)}	&	\textbf{Standard Deviation}\\
		\hline
        \textbf{Known Vectors - ARM} & & \\
        \hline
		Noisy data (no filter)	 	&	  25 	&	2.2\\
		Data filtered by NF Technique  	&	11.1 	&	0.38\\
        \hline
        \textbf{Unknown Vectors - ARM} & & \\
        \hline	
		Noisy data (no filter)	 	&	26.4 	&	2.2\\
		Data filtered by NF Technique	  	&	11.1 	&	0.41\\
        \hline
        \textbf{Known Vectors - Intel} & & \\
        \hline	
		Noisy data (no filter)	&	12.6 	&	1.2\\
		Data filtered by NF Technique	&	8.6 	&	0.18\\
        \hline
        \textbf{Unknown Vectors - Intel} & & \\
        \hline	
		Noisy data (no filter)	&	13.5 	&	0.86\\
		Data filtered by NF Technique	&	10.6 	&	0.18\\
		\hline
	\end{tabular}%
    }
	\end{center}
\end{table}

\begin{figure}[ht]
\centering
	\subfloat[ARM]
	{	\label{numnoise1a}
		\includegraphics[width=0.475\textwidth]
		{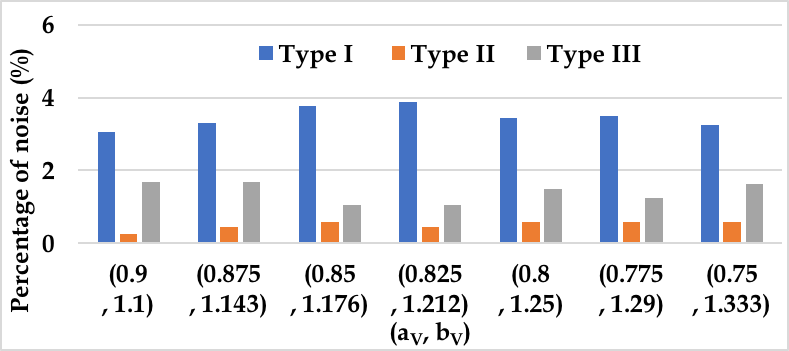}
	}
	\hfill
	\subfloat[Intel]
	{	\label{numnoise1b}
		\includegraphics[width=0.475\textwidth]
		{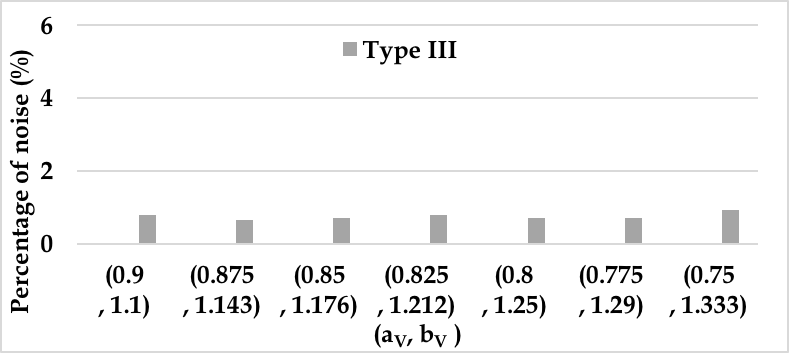}
	}
	\caption{Percentage of noise vectors before filtering}
	\label{fig:num_noise_1}
\end{figure}


\section{Design of a Two Stage Power Model}
\label{sec:twostagepowermodel}
In this section, we first explore three classic power models to understand their accuracy. This exploration motivates the need for a new power model that can work across multiple processors with low prediction errors. Thus, we design, develop and validate a Two Stage Power Model (TSPM).

\subsection{Motivation}
We evaluate three classic models (Linear Regression Power Model (LRPM), Neural Network Power Model (NNPM), and Support Vector Machine Power Model (SVMPM)) by measuring accuracy in predicting power in terms of estimation error $Error$ (defined in Section~\ref{sec:evaluationofmethodandtechnique}).

The HCS method was used to select the hardware counters as shown in Table~\ref{tab:hardwarecountersarmntree} and Table~\ref{tab:hardwarecountersintelntree}. The NF technique was then used to filter noise from the sampled vectors. The three classic power models were tested for accuracy. The training and testing strategy considered in Section~\ref{sec:evaluationofmethodandtechnique} was used. 

\begin{figure}
\centering
	\subfloat[On the ARM processor]
	{	\label{fig:comparisonclassicmodelsarm}
		\includegraphics[width=0.47\textwidth]
		{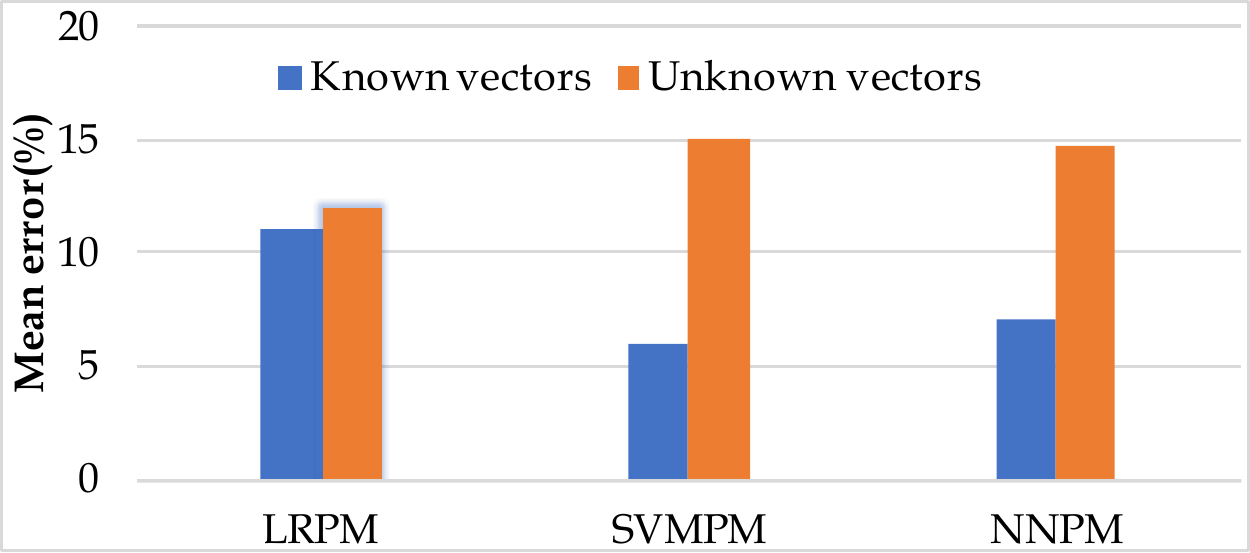}
	}
	\hfill
	\subfloat[On the Intel processor]
	{	\label{fig:comparisonclassicmodelsintel}
		\includegraphics[width=0.47\textwidth]
		{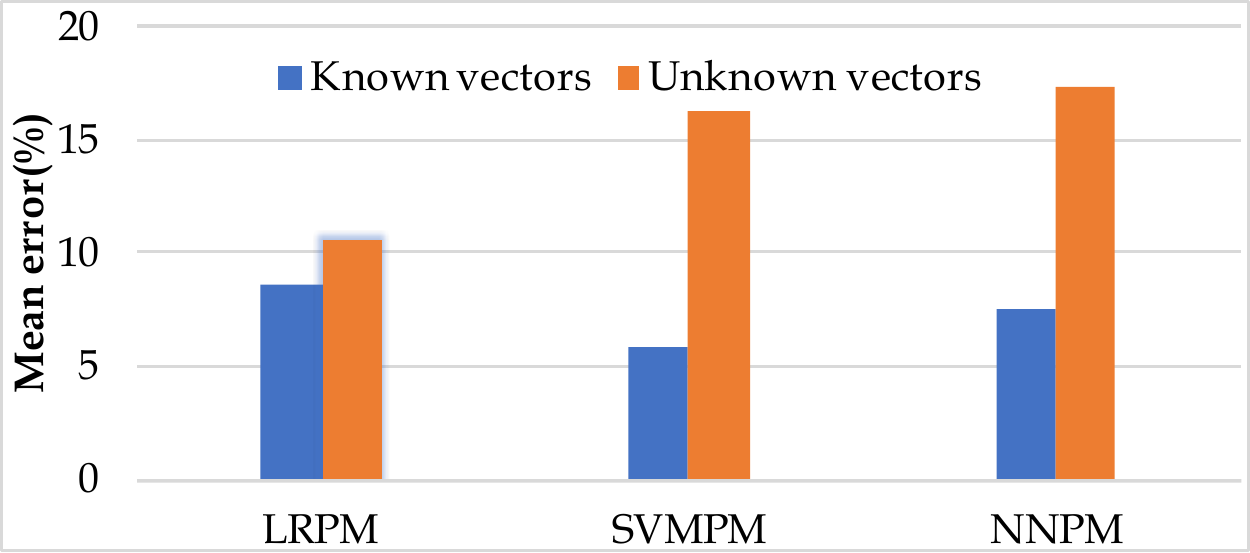}
	}\\
	\caption{Accuracy of classic power models for known and unknown vectors}
	\label{fig:comparison}
\end{figure}

Figure~\ref{fig:comparison} shows the mean error of the classic power models for predicting dynamic power when testing Known and Unknown vectors on ARM and Intel processors. On both processors SVMPM is more accurate for predicting dynamic power of Known vectors. This indicates that SVMPM fits the training data well. Compared to SVMPM, LRPM relatively under fits the data and results in lower accuracy. However, for Unknown vectors LRPM is more accurate. This is surprising, but is because more sophisticated models, such as SVMPM and NNPM, may over fit the data leading to lower accuracy for Unknown vectors than a simpler model (LRPM). 

The key observation is that there is no off-the-shelf power model that achieves accuracy of the best performing power model for known and unknown vectors. For example, SVMPM has the lowest error for predicting known vectors, but has higher error than the LRPM for predicting unknown vectors. This poses a problem in real-time power estimation - if an unknown vector is sampled because it was not used in training the model, then the model will produce inaccurate estimations. Moreover, it is time consuming to identify whether a vector is known or unknown. To make use of SVMPM for known vectors or LRPM for unknown vectors requires an additional method for identifying an incoming vector. However, this will make the model impractical for real-time use. Therefore, there is motivation for designing a new power model that can reduce the effect of over-fitted models to predict unknown vectors with higher accuracy than classic power models, but at the same time achieve low error rates for known vectors. 

\subsection{Design}
A novel power model, referred to as the Two Stage Power Model (TSPM) which takes advantage of the low variance of a simple model, the LR based power model, and of the low bias of sophisticated models, such as the SVM based difference model, is proposed.
TSPM has two stages. In the first, linear regression is used to estimate a basic power value of an incoming vector. In the second, a Support Vector Machine refines the basic value to improve estimation accuracy. TSPM operates in two phases: training and prediction. 

\subsubsection{Training of TSPM}
Algorithm \ref{trainfortspm} describes the training process of TSPM. First, an LRPM is developed using a training dataset consisting of profiled vectors (hardware counters and corresponding measured dynamic power) (Line 2). Then a difference based training dataset is constructed (Lines 3-8) by replacing the measured power of each vector in the original training set with the difference between the measured power and the prediction value estimated by the LRPM (Line 8). Finally, using the difference training set, a SVM based difference model is built (Line 9).

\begin{algorithm}[t] \small
\caption{Training process of TSPM}
\label{trainfortspm}
\begin{algorithmic}[1]
\Procedure{Train\_Model}{$training\_vectors$}
\State $LRPM \gets build\_model\left(LR, training\_vectors\right)$ 
\State $difference\_vectors \gets training\_vectors$ 
\State $n \gets sizeof\left(training\_vectors\right)$ 
\For{$i=0$ to $n-1$}
   \State $basic\_value \gets predict\left(LRPM, training\_vectors[i]\right)$
   \State $difference \gets training\_vectors[i,1] - basic\_value$
   \State $difference\_vectors[i,1] \gets difference$
\EndFor
\State $SVMDM \gets build\_model\left(SVM, \newline
difference\_vectors\right)$ 

\State \textbf{Return} $LRPM, SVMDM$
\EndProcedure

\end{algorithmic}
\end{algorithm}

\subsubsection{Prediction of TSPM}
Algorithm \ref{testfortspm} describes the prediction process of TSPM. For an incoming vector, both LRPM and SVMDM obtained from Algorithm~\ref{trainfortspm} are used for predicting. The LRPM is used to predict the basic power value (Line 2) and the SVMDM is used to estimate the difference between the measured power and the estimated power of LRPM (Line 3). We adopt a strategy to offset the basic power value with the difference, such that the final predicted power is obtained by summing the basic power and the difference (Line 4).

\begin{algorithm}[t]\small
\caption{Prediction process of TSPM}
\label{testfortspm}
\begin{algorithmic}[1]
\Procedure{Predict\_power}{$test\_vector$,$LRPM$,$SVMDM$}
\State $basic\_value \gets predict\left(LRPM, test\_vector\right)$
\State $difference \gets predict\left(SVMPM, test\_vector\right)$
\State $power \gets basic\_value + difference$
\State \textbf{Return} $power$
\EndProcedure
\end{algorithmic}
\end{algorithm}

\subsection{Comparing TSPM and Classic Power Models}
In this section, the accuracy of the proposed TSPM against classic power models is evaluated. The vectors and the training-testing strategy presented in Section~\ref{sec:evaluationofmethodandtechnique} are used.

\begin{figure*}
\centering
	\subfloat[On ARM]
	{	\label{resunknownarm}
		\includegraphics[width=0.485\textwidth]
		{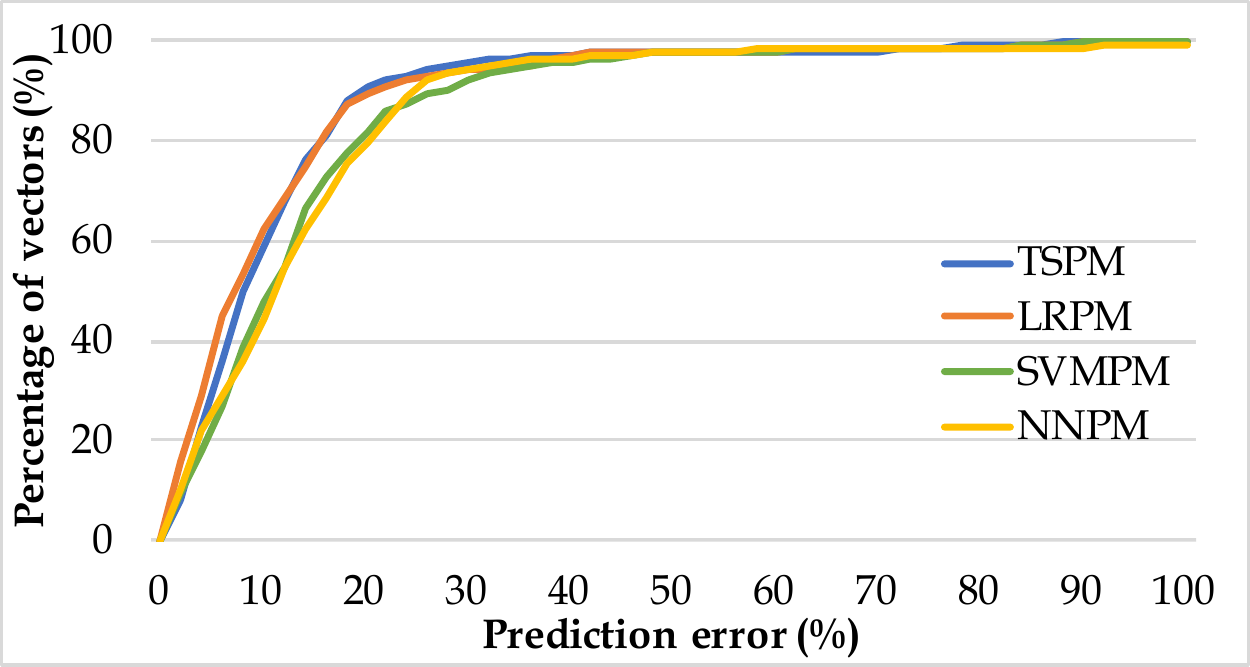}
	}
	\hfill
	\subfloat[On Intel]
	{	\label{resunknownintel}
		\includegraphics[width=0.485\textwidth]
		{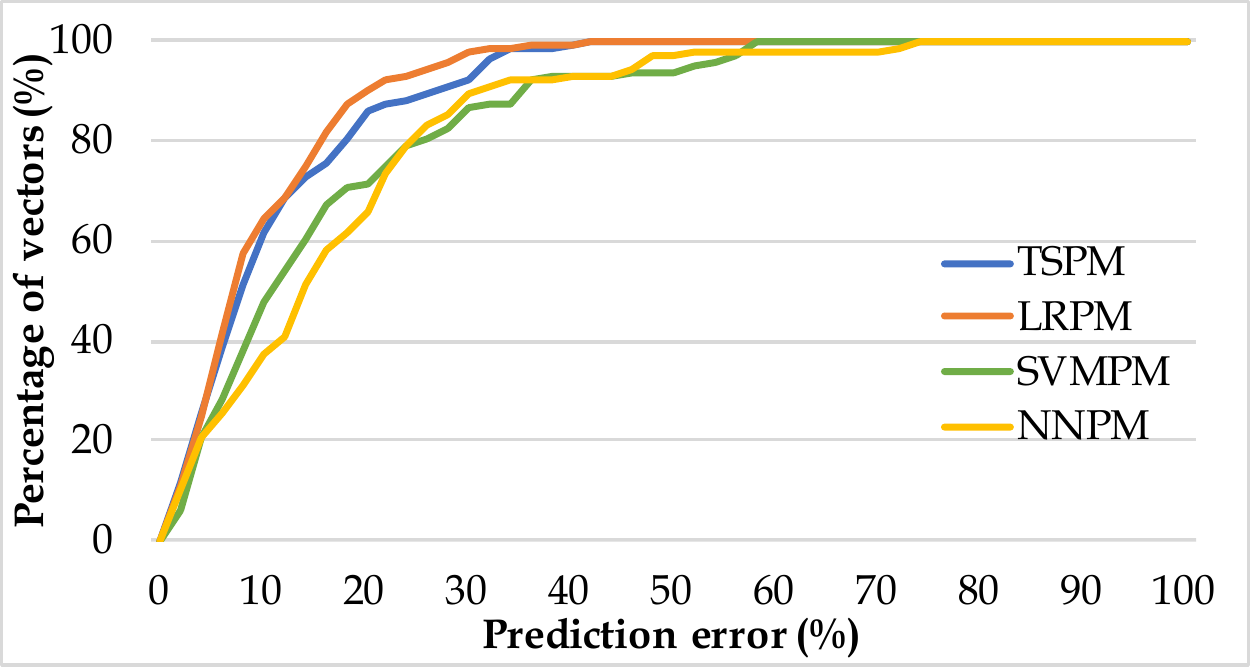}
	}\\
	\caption{Prediction accuracy of different models for Unknown vectors}
	\label{res}
\end{figure*}

Figure~\ref{res} shows the prediction accuracy of TSPM compared to classic power models for unknown vectors on ARM and Intel. TSPM obtains accuracy similar to the best classic model (LRPM). Nearly 60\% of unknown vectors can be predicted with error less than 10\% using TSPM and LRPM.

\begin{table}[ht]
	\caption{Percentage mean error of different power models}
	\label{tab:meanerrorall}
\begin{center}
\centering    
   \resizebox{0.48\textwidth}{!}{%
	\begin{tabular}{c c c c c c}
		\hline	       		
		& &	\textbf{TSPM}	&	\textbf{LRPM} & \textbf{SVMPM} & \textbf{NNPM}\\
		\hline
        \multirow{2}{*}{\textbf{Known}} & ARM & 6.6 & 11.1 & 6 & 7.1 \\
        \cline{2-6}
        & Intel & 6.3 & 8.6 & 5.8 & 7.5\\
        \hline
        \multirow{2}{*}{\textbf{Unknown}} & ARM & 11.9 & 11.1 & 15 & 14.8\\
        \cline{2-6}
        & Intel & 11.5 & 10.6 & 16.3 & 17.4\\
        \hline 		
        \multirow{2}{*}{\textbf{All}} & ARM & \textbf{7.9} & 11.1 & 8.3 & 9.0\\
        \cline{2-6}
        & Intel & \textbf{7.6} & 9.1 & 8.4 & 10.0\\
        \hline 
	\end{tabular}%
    }
	\end{center}
\end{table}

Table~\ref{tab:meanerrorall} summarises the evaluation of TSPM against the classic power models. There is no single classic power model that performs well on both known and unknown vectors. However, TSPM has a lower mean error rate for predicting all vectors, including known and unknown vectors when compared to all three classic models. The key result is that it is not possible to use different models for known and unknown vectors since it is time consuming to identify whether a vector is a known/unknown in real-time. The TSPM model can be employed to address this problem since it has better accuracy (7.9\% and 7.6\% error on ARM and Intel) for all vectors when compared to the classic models. It is also noted that TSPM works across processor architectures. 

We measured the computation time of LRPM and SVMDM. 
For an incoming vector, the wall clock time taken for estimating the basic value using LRPM is approximately equal to 1.1ms and time for estimating the difference value using SVMDM is approximately 0.95ms. The total time taken for estimating power of each vector using TSPM is approximately equal to 2.05ms. Compared to the mean sampling time interval (1 second), the overhead of TSPM is negligible.

\section{Related Work}
\label{sec:relatedwork}
Research in power modelling has led to (i) instruction-level, (ii) coarse-grain utilisation, and (iii) hardware counter-based models. These models are typically platform dependent and work across homogeneous processors in data centers. 

Instruction-level models are based on estimating power at the software level~\cite{instructionlevelmodel-1}. Typically, the number of instructions is used to estimate the overall power of an application that is executed. The power of an application is estimated by aggregating the power of all individual instructions and inter-instruction effects~\cite{instructionlevelmodel-2}. This requires extensive knowledge of the entire instruction set. It is cumbersome to obtain power of each instruction and the overhead of all instruction pairs, thereby rendering these models impractical for real use.   

Coarse-grained utilisation-based power models estimate power as a function of utilisation of individual components of a system, such as CPU and memory. The relation between CPU utilisation and power of single-core and dual-core processors using a quadratic function and a linear function, respectively is known~\cite{utilisationlevelmodel-1}. Other research estimates energy as the product of the overall energy of the core and the core utilisation of a task
~\cite{utilisationlevelmodel-2}. Although these models are easy to implement, they are not accurate since power depends not only on utilisation, but also on the type of operation. For example, floating point operations require more power than integer operations.   

Hardware counter-based power models may be viewed as fine-grained utilisation models. The overall system power is the sum of power of sub-components. Each sub-component's power is modelled as a function of hardware counters, but the question is which counters to use. For example, research demonstrates: (a) the development of a power proxy by selecting 50 activity counts from hundreds of candidates~\cite{lrpm-3}, (b) the definition of 22 physical power sub-units for an Intel processor and the use of 24 event metrics for modelling all sub-units~\cite{lrpm-isci-1}, and (c) the definition of 8 power components for an Intel processor and the selection of 15 hardware counters to determine the activity of these components~\cite{lrpm-2}.
These models are relatively simpler than instruction-level models, but are also more accurate than coarse-grained utilisation models. 


\section{Conclusions}
\label{sec:conclusions}



This paper proposed power modelling techniques focused on automation, simplicity and high accuracy. To achieve this, an automated hardware counter selection method that selects hardware counters relevant to power for both ARM and Intel processors is developed. In current research this is manual and extensively explores all hardware counters obtained from a processor. This is not feasible as more diverse processors are added to the computing ecosystem. The accuracy of power estimation is improved by up to 15\% by using the automated selection method. Secondly, a noise filter based on clustering that can reduce the mean error in the power modelling data by up to 55\% was developed. Finally, a two stage power model that surmounts the challenges in using existing power models across multiple architectures was designed. It was demonstrated that this model predicts dynamic power with less than 8\% error on both ARM and Intel processors, which is an improvement over classic power models. 

In the future, we aim to improve the accuracy by developing: (i) benchmarking techniques that capture a program's power consumption, and (ii) time-series based dynamic calibration to improve the estimation accuracy for Unknown vectors.


\section*{Acknowledgment}
Dr Blesson Varghese is supported by a Royal Society Short Industry Fellowship and is funded by Rakuten Mobile, Japan. 

\bibliographystyle{IEEEtran}
\bibliography{references}

\end{document}